\documentclass{article}
\usepackage{spconf,amsmath,graphicx}
\usepackage{amsfonts}
\usepackage{multirow}
\usepackage{hyperref}
\usepackage{url,times,bm,siunitx,float}
\usepackage{color}
\usepackage{balance}

\title{
DIFFERENTIABLE CONSISTENCY CONSTRAINTS
\\FOR IMPROVED DEEP SPEECH ENHANCEMENT
}
\name{
\parbox{\linewidth}{
\centering
Scott Wisdom,
John R.\ Hershey,
Kevin Wilson,
Jeremy Thorpe,\\
Michael Chinen,
Brian Patton,
Rif A.\ Saurous
}}
\address{Google Research}
\begin{document}
\ninept
\maketitle
\begin{abstract}
In recent years, deep networks have led to dramatic improvements in speech enhancement by framing it as a data-driven pattern recognition problem. In many modern enhancement systems, large amounts of data are used to train a deep network to estimate masks for complex-valued short-time Fourier transforms (STFTs) to suppress noise and preserve speech. However, current masking approaches often neglect two important constraints: STFT consistency and mixture consistency. Without STFT consistency, the system's output is not necessarily the STFT of a time-domain signal, and without mixture consistency, the sum of the estimated sources does not necessarily equal the input mixture. Furthermore, the only previous approaches that apply mixture consistency use real-valued masks; mixture consistency has been ignored for complex-valued masks.

In this paper, we show that STFT consistency and mixture consistency can be jointly imposed by adding simple differentiable projection layers to the enhancement network. These layers are compatible with real or complex-valued masks. Using both of these constraints with complex-valued masks provides a 0.7 dB increase in scale-invariant signal-to-distortion ratio (SI-SDR) on a large dataset of speech corrupted by a wide variety of nonstationary noise across a range of input SNRs.
\end{abstract}
\section{Introduction}

In recent years, deep neural networks (DNNs) have led to dramatic improvements in speech enhancement.  Typically deep networks are trained to estimate masks for complex-valued short-time Fourier transforms (STFTs) to suppress noise and preserve speech. 
That is, given $N$ time-domain samples of an input mixture of $J$ sources:
\begin{equation}
    {\bf y} = \sum_{j=1}^J {\bf x}_j
    \label{eq:mix},
\end{equation}
a DNN is trained to produce a mask 
${\bf M}_j$
for each source given the mixture. For each of the $J$ sources, the mask is applied to the mixture STFT 
${\bf Y}=\mathcal{S}\{{\bf y}\}$ of shape $F\times T$, 
and the separated audio signal is recovered by applying the inverse STFT operator:
\begin{equation}
    \hat{\bf x}_j
    =
    \mathcal{S}^{-1}
    \big\{
    {\bf M}_j
    \odot
    {\bf Y}
    \big\},
    \label{eq:seperated}
\end{equation}
where $\mathcal{S}$ and $\mathcal{S}^{-1}$ are forward and inverse STFT operators.%

Such masking-based DNN approaches have been very successful \cite{narayanan2013ideal, erdogan2015phase, williamson2016complex, wilson2018exploring}. However, existing approaches have two deficiencies. First, the loss function used to train the enhancement network is typically measured on the masked noisy STFT. The problem with this is that applying an arbitrary mask does not produce a consistent STFT, in the sense that the masked STFT could not be computed from any real-valued time-domain signal. Therefore, if the masked STFT is inverted then recomputed, the resulting magnitudes and phases will be different. Second, some approaches that use a real-valued mask and all approaches that use a complex-valued mask do not leverage the basic assumption given by the model (\ref{eq:mix}): that the separated sources should add up to the original mixture signal. 

In this paper, we show that STFT and mixture consistency can be enforced by adding simple end-to-end layers in the enhancement network. These two techniques can be combined with any masking-based speech enhancement model to improve performance. Also, these constraints are compatible with complex-valued masks, which for the first time allows systems that use complex-valued masks to take advantage of the basic assumption of mixture consistency.

\section{Relation to prior work}

STFT consistency has been exploited by a number of works \cite{LeRoux2008SAPA09b, LeRoux2010DAFx09, LeRoux2010LVA09, Magron2017WASPAA10, leroux2013consistent}, but usually in a model-based context that requires iterative algorithms.
In contrast to our work, none of these approaches have combined STFT consistency with DNNs. Since we use a single differentiable consistency constraint layer within a DNN, we do not require iterations to enforce consistency.

Gunawan and Sen \cite{gunawan2010iterative} proposed adding mixture consistency constraints to improve Griffin-Lim, resulting in the multiple input spectrogram inversion (MISI) algorithm. Wang et al.\ \cite{wang2018end} trained a real-valued masking-based system for speech separation through unfolded MISI iterations. Strumel and Daudet \cite{sturmel2012iterative} proposed iterative partitioned phase reconstruction (PPR), which also relied on mixture consistency. In contrast to these works, we do not require multiple iterations for phase estimation, and furthermore we experiment with estimating phase using a complex-valued mask.

Lee et al.\ \cite{lee2018phase} recently used a soft squared error penalty between the warped magnitudes of the true and estimated mixture STFTs to promote mixture consistency. In contrast, we use an exact projection step implemented as a neural network layer.

\begin{figure*}[!ht]
    \centering
    \includegraphics[width=0.675\linewidth]{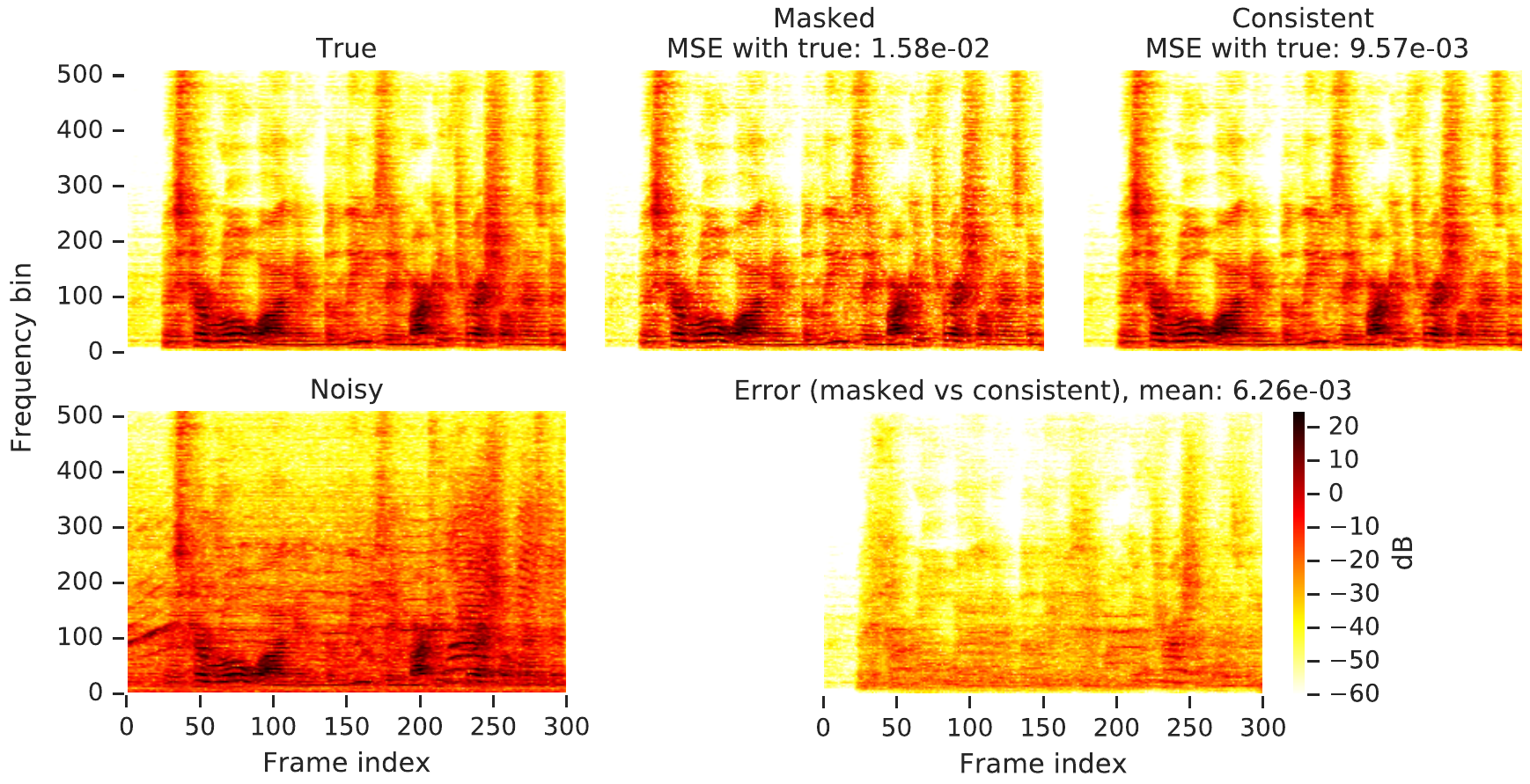}
    \vspace{-10pt}
    \caption{Illustration of the effect of STFT consistency. A noisy STFT (spectrogram in lower left) is masked using an ideal phase-sensitive mask (\ref{eq:ideal_psm}) to recover a target speech signal (spectrogram in upper left). Then, this masked STFT is made consistent using the projection (\ref{eq:stftproj}). Notice that the masked spectrogram (upper middle panel) is substantially different from the STFT-consistent spectrogram (upper right panel), with the magnitude-squared error visualized in the lower right panel.}
    \label{fig:stft_consistency}
    \vspace{-10pt}
\end{figure*}

\section{Consistency}
\subsection{STFT consistency}

When a STFT uses overlapping frames, applying a mask to the STFT of a mixture signal, whether the mask is real or complex-valued, does not necessarily produce a consistent STFT \cite{LeRoux2008SAPA09b}. A consistent STFT ${\bf X}$ is one such that there exists a real-valued time-domain signal ${\bf x}$ satisfying ${\bf X}=\mathcal{S}\{{\bf x}\}$. 
A STFT ${\bf X}$ is consistent if it satisfies
\begin{equation}
    {\bf X}
    =
    \mathcal{S}
    \Big\{
        \mathcal{S}^{-1}
            \big\{
                {\bf X}
            \big\}
    \Big\}
    =
    \mathcal{P}_S
    \{
        {\bf X}
    \},
    \label{eq:stftproj}
\end{equation}
where 
$\mathcal{P}_S$ 
refers to the projection performed by the sequence of inverse and forward STFT operations.

An important reason to enforce consistency on separated estimates ${\bf M}_j\odot{\bf Y}$ is that these estimates are used in loss functions and metrics used during deep network training. If the masked STFT is simply used in a loss function, e.g.\ mean-squared error between magnitudes of the masked and ground-truth spectrograms,
the magnitudes 
$|\hat{\bf X}_j|=|{\bf M}_j\odot{\bf Y}|$ 
do not necessarily correspond to the STFT magnitudes of the reconstructed time-domain signal, $|\mathcal{S}\{\hat{\bf x}_j\}|$. Thus, the loss function will not be accurately measuring the spectral magnitude of the estimate.

\vspace{-7pt}
\subsubsection{Illustration of STFT consistency}

Figure \ref{fig:stft_consistency} illustrates the effect of STFT consistency. A clean speech example ${\bf s}$ (spectrogram in upper left panel) is embedded in nonstationary noise ${\bf v}$ at 8 dB SNR (mixture spectrogram in lower left panel). Then, we compute an oracle phase-sensitive mask \cite{erdogan2015phase} as
\begin{equation}
    \frac{|S_{f,t}|}
    {|S_{f,t} + V_{f,t}|}
    \cdot\cos(\angle S_{f,t} - \angle Y_{f,t}).
    \label{eq:ideal_psm}
\end{equation}
This mask is applied to the noisy STFT to create a masked STFT (spectrogram shown in upper middle panel). Then, we use the projection (\ref{eq:stftproj}) to create a consistent STFT of the enhanced speech (spectrogram in upper right panel). The lower right panel visualizes the magnitude-squared error between the masked and consistent STFTs.

Note that the masked STFT (upper middle panel) differs substantially from the consistent STFT (upper right panel). This indicates the importance of STFT consistency: if a neural network training loss is measured on the masked magnitude instead of the consistent magnitude, then the loss is not looking at the actual spectrogram of the enhanced time-domain signal.
In particular, the magnitude-squared error between consistent and true STFTs (9.57e-3) is less than the magnitude-squared error between masked and true STFTs (1.58e-2). This discrepancy emphasizes the importance of having the network compute training losses on consistent STFTs, since otherwise the network is focusing on irrelevant parts of the error.

\vspace{-5pt}
\subsubsection{Backpropagating through STFTs}

A natural way to enforce STFT consistency is to simply project estimated signals using the constraint (\ref{eq:stftproj}). Since the forward and inverse STFT operations are linear transforms implemented in TensorFlow\footnote{\url{https://www.tensorflow.org/api_docs/python/tf/contrib/signal/inverse_stft}}, this projection can simply be treated as an extra layer in the network.

\vspace{-5pt}
\subsection{Mixture consistency}

An obvious constraint on separated signals comes from the original mixing model (\ref{eq:mix}), as it is natural to assume that estimates of the separated signals should add up to the original mixture. This is equivalent to the complex STFTs of the estimated sources adding up to the complex mixture STFT:
\begin{equation}
    \sum_{j} \underline{X}_{j, f, t} = Y_{f,t} \; \; \forall t, f,
\end{equation}
where ${\underline{X}}_{j,f,t}$ is the mixture-consistent TF bin for source $j$.

In the past, this mixture consistency constraint has been enforced using real-valued masks that are made to sum to one across the sources, which is equivalent to using the projection (\ref{eq:mixproj}).
However, when the masks are complex-valued, constraining these masks to sum to one across sources is too restrictive, since complex masks might need to modify the phase differently for different sources in the same time-frequency (TF) bin. 

In this section, we describe a simple differentiable mixture projection layer that can enforce mixture consistency for any type of masking-based enhancement or separation method, including real-valued and complex-valued masks and explicit phase prediction.

\vspace{-5pt}
\subsubsection{Backpropagating through mixture-consistent projection}

To ensure mixture consistency without putting explicit constraints on the masks, we project the masked estimates to the nearest points on the subspace of mixture-consistent estimates. To do this, we solve the following optimization problem for each TF bin:
\begin{equation}
    \begin{aligned}
        & \underset{\underline{\bf X}_{f,t}\in\mathbb{C}^J}{\text{minimize}}
        & & \frac{1}{2} \sum_j |\underline{X}_{j, f, t} - \hat{X}_{j,f,t}|^2
        \\ & \text{subject to}
        & & \sum_j \underline{X}_{j,f,t} = Y_{f,t}.
    \end{aligned}
    \label{eq:mixprob}
\end{equation}
Using the method of Lagrange multipliers
and defining the estimated mixture $\hat{Y}_{f,t}=\sum_j\hat{X}_{j,f,t}$ yields the following update, which is a simple projection that can be added as a layer in the network and backpropagated through:
\begin{equation}
    \underline{X}_{j,f,t}
    =
    \hat{X}_{j,f,t}
    +
    \frac{1}{J}
    (
        Y_{f,t} - \hat{Y}_{f,t}
    ).
    \label{eq:mixproj}
\end{equation}

\subsubsection{Weighted mixture-consistent projection with uncertainty}

If we have {\it a priori} knowledge of the uncertainty of the source estimates, a weighted version of the problem (\ref{eq:mixprob}) can be used to project the sources. Assume that in each frequency bin, each estimated source can be modeled as a zero-mean complex-valued circular Gaussian with variance $v_{j, f, t}$. 

Given variances $v_{j,f,t}$, the weighted problem is
\begin{equation}
    \begin{aligned}
        & \underset{\underline{\bf X}_{f,t}\in\mathbb{C}^J}{\text{minimize}}
        & & 
        \frac{1}{2} 
        \sum_j
        \frac{1}{v_{j,f,t}}
        \big|\underline{X}_{j,f,t} - \hat{X}_{j,f,t}\big|^2
        \\
        & \text{subject to}
        & & \sum_j \underline{X}_{j,f,t} = Y_{f,t}.
    \end{aligned}
\end{equation}
Using the method of Lagrange multipliers yields
\begin{equation}
    \underline{X}_{j,f,t}
    =
    \hat{X}_{j,f,t}
    +
    \frac{v_{j,f,t}}{\sum_{j'} v_{j',f,t}}
    (
        Y_{f,t} - \hat{Y}_{f,t}
    ).
    \label{eq:wmixproj}
\end{equation}
Notice that if all sources have equal uncertainty, i.e. that $v_{j,f,t}=v_{j',f,t}$ for all $j\not = j'$, then the weighted projection (\ref{eq:wmixproj}) is equivalent to the unweighted projection (\ref{eq:mixproj}).

\begin{figure}
    \centering
    \includegraphics[width=0.6\linewidth]{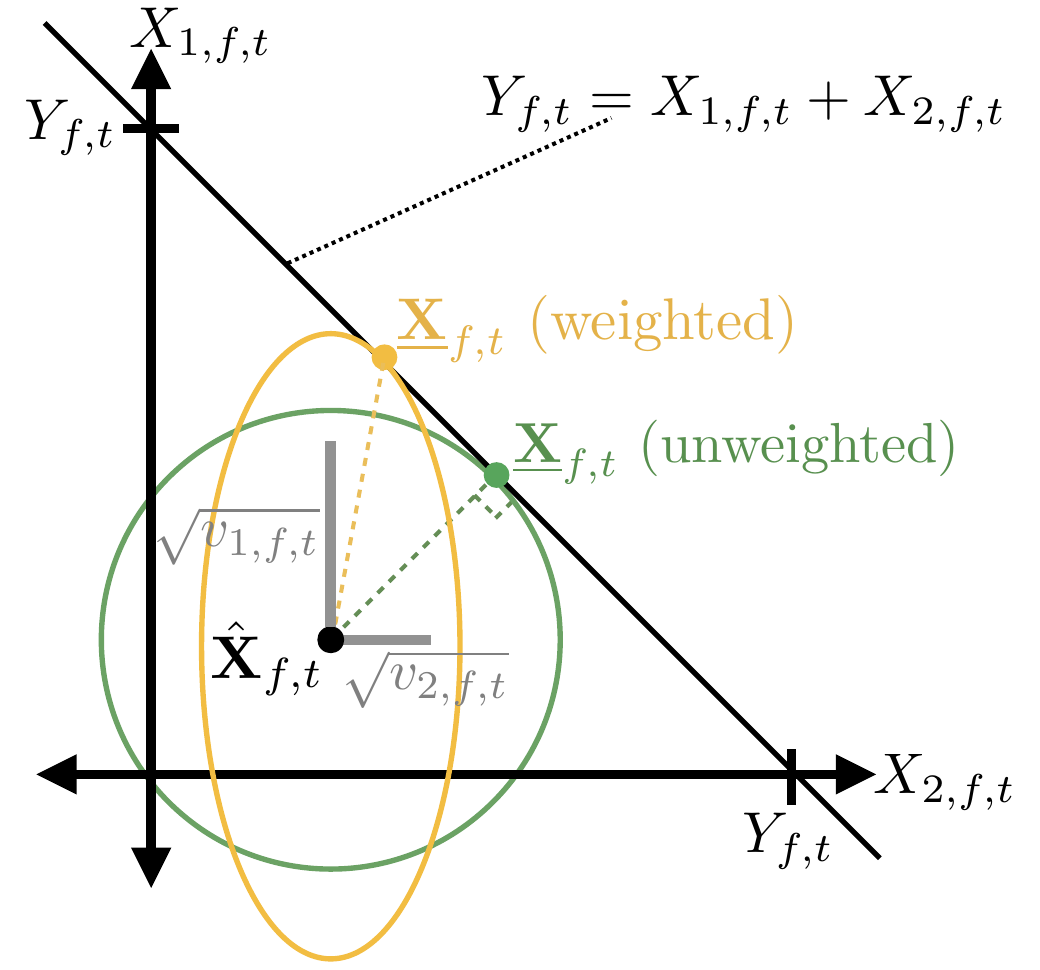}
    \vspace{-7.5pt}
    \caption{Geometric illustration of mixture consistency for two real-valued sources in a single time-frequency bin.}
    \label{fig:mix_illus}
    \vspace{-10pt}
\end{figure}

Figure \ref{fig:mix_illus} illustrates unweighted and weighted mixture consistency for two sources. For visualization, we depict the sources' TF bins $X_{1, f, t}$ and $X_{2, f, t}$ as real-valued instead of complex-valued. Note that the projected mixture-consistent estimate $\underline{\bf X}_{f,t}$ is given by the intersection of the mixture constraint line and the ellipse representing contours of a diagonal-covariance Gaussian pdf. When the variances $v_{j,f,t}$ are the same for all sources, the ellipse becomes a circle, and the mixture-consistent projection is orthogonal to the constraint surface.

\subsection{Order of consistency operations}

The STFT and mixture consistency operations (\ref{eq:stftproj}) and (\ref{eq:mixproj}) are linear projections. Though linear projections do not in general commute, the order in which these projections are applied does not matter. As a proof, we start from the case where STFT consistency (\ref{eq:stftproj}) is applied after mixture consistency (\ref{eq:mixproj}). Notice that since $\mathcal{P}_S$ is linear, and since a linear combination of consistent STFTs is still a consistent STFT, we can write this as being equivalent to applying mixture consistency after STFT consistency:
\begin{equation}
\begin{aligned}
    \underline{\bf X}_{j}
    =&
    \mathcal{P}_S
    \{
        \hat{\bf X}_j
        + \frac{1}{J}({\bf Y} - \sum_{j'} \hat{\bf X}_{j'})
    \}
    \\
    \Leftrightarrow
    \underline{\bf X}_{j}
    =&
    \mathcal{P}_S
    \{
        \hat{\bf X}_j
    \}
    + \frac{1}{J}
        {\bf Y}
    - \frac{1}{J}
    \sum_{j'}
    \mathcal{P}_S
    \{
        \hat{\bf X}_{j'}
    \}.
\end{aligned}
\label{eq:order_proof}
\end{equation}

However, weighted mixture projections that apply a different weight per TF bin are not orthogonal to STFT consistency projection, since the relation (\ref{eq:order_proof}) relies on $1/J$ not depending on time or frequency. Despite this lack of commutativity, we do still observe benefits of combining STFT consistency with weighted mixture consistency. Jointly imposing STFT and weighted mixture consistency is more complicated, which we defer to future work.

\begin{figure*}[!ht]
  \centering
  \centerline{\includegraphics[width=0.9\textwidth]{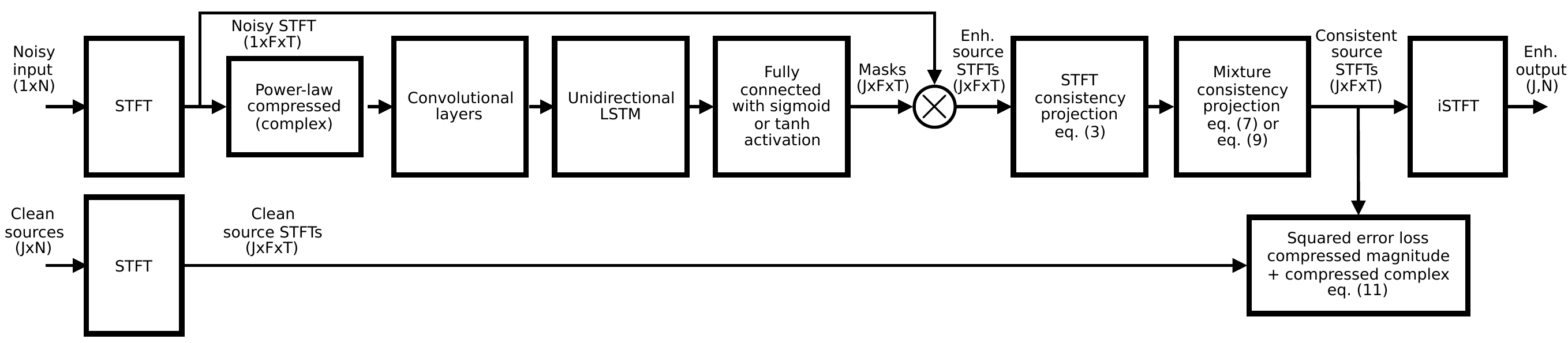}}
  \vspace{-5pt}
  \caption{System architecture when both STFT consistency and mixture consistency are used.
  }
  \label{fig:architecture_block_diagram}
\end{figure*}

\begin{figure*}[!ht]
    \centering
    $\vcenter{\hbox{\includegraphics[width=0.83\linewidth]{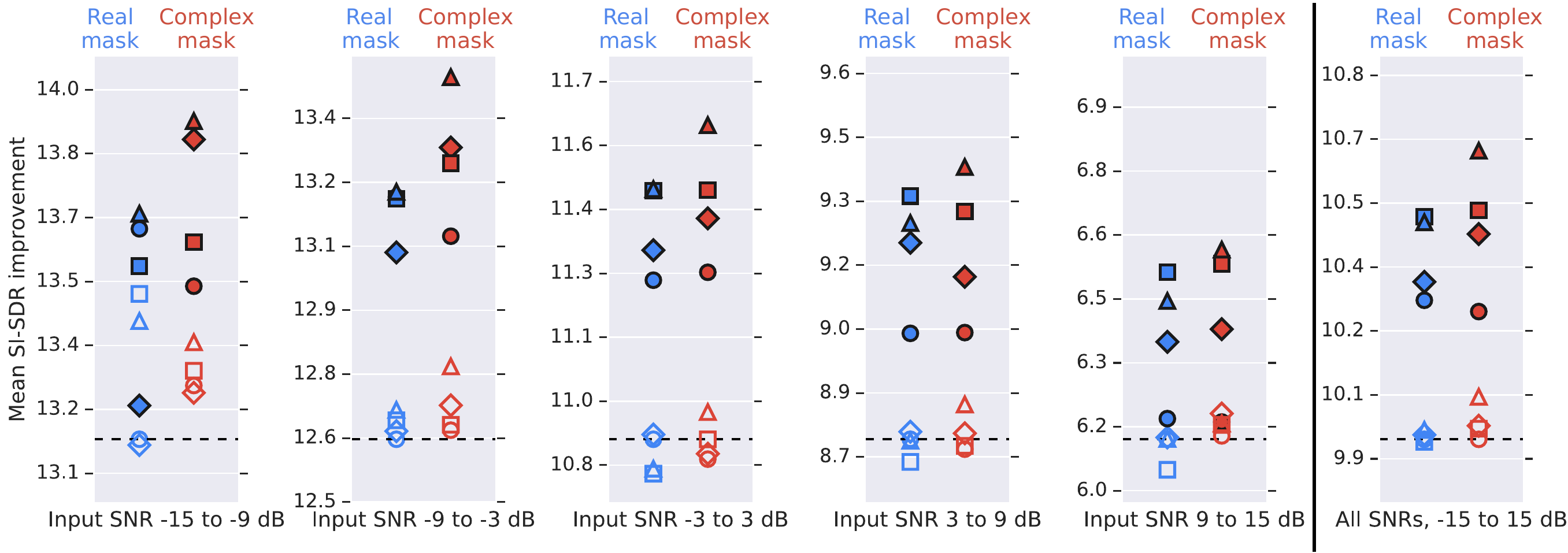}}}$
    $\vcenter{\hbox{\includegraphics[width=0.13\linewidth]{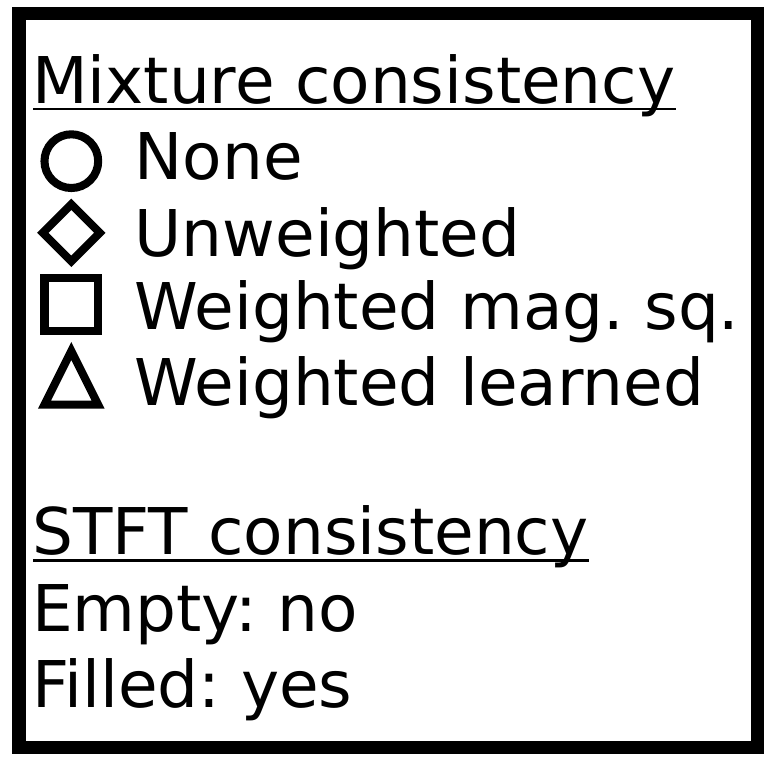}}}$
    \caption{Mean improvement in SI-SDR for different input SNRs for the the test set. Notice that using both types of consistency almost always improves performance, for both real-valued and complex-valued masks. The dashed lines indicate the performance of the baseline model which uses a real-valued mask with noisy phase and neither consistency constraint (blue empty circle).}
    \label{fig:results}
\end{figure*}

\section{Experiments}

\subsection{Dataset}

We use a dataset constructed from publicly-available data. Speech is taken from a subset of LibriSpeech \cite{panayotov2015librispeech}, where train, validation, and test sets use nonoverlapping sets of speakers, and audio has been filtered using WADA-SNR \cite{kim2008robust} to ensure low levels of background noise. Noise data is sourced from \url{freesound.org}, which contains a large variety of music, musical instruments, field recordings, and sound effects. Files that are very long, have a large number of zero samples, or exhibit clipping are filtered out.
The sampling rate is 16kHz.

Speech and noise are mixed with normally-distributed SNRs, with a mean of 5 dB and standard deviation of 10 dB. We also apply a random gain to the mixture signals, with a mean of -10 dB and a standard deviation of 5 dB. These random gain parameters were chosen such that the clipping is minimal in the audio. After mixing, the duration of the training set is 134.2 hours, and the validation and test sets are  6.2 hours each.

\subsection{Model architecture and training}
Our model, shown in Figure \ref{fig:architecture_block_diagram}, uses a modified version of a similar architecture that recently achieved state-of-the-art performance on the CHiME2 dataset \cite{wilson2018exploring}. The model consists of a convolutional front-end operating on spectral input features, a single unidirectional LSTM with residual connections and width of 400, and two fully-connected layers with 600 hidden units each. The input features are power-compressed STFTs, computed as $X^{0.3}:=|X|^{0.3}e^{j\angle X}$. STFTs are computed using 50 ms Hann windows with a 10 ms hop and FFT length of 1024.

The training loss, $L$, for all networks is as follows:
\begin{equation}
\begin{aligned}
  \label{eqn:loss}
  L  = \sum_{j=1}^2 z_j \sum_{f,t} & 
  \left[
  \big(
    |X_{j,f,t}|^{0.3} - |\hat{X}_{j,f,t}|^{0.3}
  \big)^2
  \right.
  \\&
  \left.
  + 0.2 \cdot
  \big|
    X_{j,f,t}^{0.3} - \hat{X}_{j,f,t}^{0.3}
  \big|^2
  \right],
\end{aligned}
\end{equation}
where $z_1=0.8$ is the speech loss weight and $z_2=0.2$ is the noise loss weight. We train and validate on fixed-length, three-second clips. The Adam optimizer \cite{kingma2014adam} is used with a batch size of 8, learning rate of $3\times 10^{-5}$, and default parameters otherwise. 

Our proposed consistency constraints are compatible with both real and complex-valued masks. For real-valued masking, the network predicts a single scalar value through a sigmoid nonlinearity for each TF bin, and the noisy phase is used for reconstruction.
To perform complex-valued masking, we use an approach similar to that of Williamson et al.\ \cite{williamson2016complex}. For each TF bin, the network predicts the real and imaginary parts of a complex-valued mask through a hyperbolic tangent (tanh) nonlinearity. This mask is multiplied with the complex noisy STFT, then reconstructed.

To implement weighted mixture consistency (\ref{eq:wmixproj}), we consider two types of weighting schemes:
\begin{enumerate}
    \item Weights $v_{j,f,t}$ are the squared estimated source magnitudes, $|\hat{X}_{j,f,t}|^2$. This has the advantage of not adding any correction signal to TF bins that have a low magnitude, which helps when there is only one signal active within a TF bin.
    \item Weights are learned by the network. The network outputs a scalar for each time, frequency, and source. A sigmoid nonlinearity is applied, producing weights $w_{1,f,t}$ for the speech source, and $w_{2,f,t}=1-w_{1,f,t}$ for the noise source.
    The terms $v_{j,f,t} / (\sum_{j'} v_{j',f,t})$ in (\ref{eq:wmixproj}) are replaced by $w_{j,f,t}$.
\end{enumerate}

\subsection{Results}

Results are shown in Figure \ref{fig:results} in terms of scale-invariant SDR (SI-SDR) improvement. SI-SDR measures  signal fidelity with respect to a %
reference signal while allowing for a gain mismatch \cite{isik2016single, leroux2018sdr}:
\begin{equation}
    \text{SI-SDR} = 10 \log_{10} 
    \frac
    {\| \alpha {\bf x}\|^2}
    {\| \alpha {\bf x} - \hat{\bf x}\|^2},
\end{equation}
where $\alpha = \textrm{argmin}_{a} \| a {\bf x} - \hat{\bf x}\|^2 = {\bf x}^T \hat{\bf x} /\|{\bf x}\|^2$.

These results are grouped into five bins based on input SNR, from -15 dB to 15 dB with bin width of 6 dB. Models that use both STFT and mixture consistency constraints almost always outperform models that do not use these constraints. The best models tend to use weighted mixture consistency with learned weights. Phase prediction provides a slight improvement in performance, especially when using a complex mask for phase prediction at lower input SNRs. The system with the best overall mean SI-SDR improvement of 10.6 dB uses complex masking, STFT consistency, and weighted mixture consistency with learned weights. This improves 0.7 dB over the baseline system, which has mean SI-SDR improvement of 9.9 dB.

\section{Conclusion}

In this paper, we have shown that the simple addition of differentiable neural network layers can be used to enforce STFT and mixture consistency on source estimates of an audio source separation network. Adding these consistency constraint layers improves speech enhancement performance on a dataset using a wide variety of nonstationary noise and SNR levels. These constraints are also compatible any type of STFT-based enhancement systems, including those that use complex-valued masks. In future work, we will combine these constraints with other types of phase estimation and generative models, as well as use them for general audio source separation, rather than just speech enhancement. Another interesting direction of future work is defining a weighted version of STFT consistency and joint weighted STFT and mixture consistency, both of which have the potential to further improve performance.

\vfill\pagebreak

\bibliographystyle{IEEEtran_nourl}
\balance
\bibliography{refs}

\end{document}